\documentclass[prl,aps,superscriptaddress,epsf,showpacs,twocolumn]{revtex4-1}
\usepackage{times}
\usepackage{graphicx}
\usepackage{float}
\usepackage{latexsym,amsmath,amssymb,bm,euscript}
\usepackage{color}
\usepackage{subfigure}
\usepackage{epstopdf}
\usepackage[colorlinks=true,linkcolor=blue,citecolor=blue,urlcolor=blue]{hyperref}
\usepackage{hyperref}
\usepackage{ulem}
\usepackage[dvipsnames]{xcolor}
\usepackage{dcolumn}     
\usepackage{slashbox}     
\usepackage{bbm}

\newcommand{\be}[0]{\begin{equation}}
\newcommand{\ee}[0]{\end{equation}}
\newcommand{\ba}[0]{\begin{eqnarray}}
\newcommand{\ea}[0]{\end{eqnarray}}
\newcommand{\mx}[0]{\begin{pmatrix}}
\newcommand{\ex}[0]{\end{pmatrix}}

\newcommand{\up}[0]{\uparrow }
\newcommand{\dn}[0]{\downarrow }

\begin{document}

\title{Orbital-selective Mott phase as a dehybridization fixed point}

\author{Haoyu Hu}
\email{hh25@rice.edu}
\affiliation{Department of Physics and Astronomy, Rice Center for Quantum Materials,  Rice University,
Houston, Texas, 77005, USA}

\author{Lei Chen}
\affiliation{Department of Physics and Astronomy, Rice Center for Quantum Materials,  Rice University,
Houston, Texas, 77005, USA}

\author{Jian-Xin Zhu}
\affiliation{Theoretical Division and Center for Integrated Nanotechnologies, Los Alamos National Laboratory, Los Alamos, New Mexico 87545, USA}

\author{Rong Yu}
\affiliation{Department of Physics and Beijing Key Laboratory of Opto-electronic Functional Materials and Micro-nano Devices,
Renmin University of China, Beijing 100872, China}

\author{Qimiao Si}
\email{qmsi@rice.edu}
\affiliation{Department of Physics and Astronomy, Rice Center for Quantum Materials,  Rice University,
Houston, Texas, 77005, USA}

\begin{abstract}
Studies on the iron-based superconductors and related strongly correlated systems
have focused attention on bad-metal normal  state in proximity
to antiferromagnetic order. 
An orbital-selective Mott phase (OSMP) has been extensively
discussed as anchoring the orbital-selective correlation 
phenomena in this regime.
Motivated by recent experiments,
we advance the notion that an OSMP is 
 synonymous to correlation-driven dehybridization.
 This idea is developed in terms of a competition between inter-orbital hopping
 and dynamical spatial spin correlations.
Within effective models that arise from extended dynamical mean-field theory (EDMFT), 
and using a combination of continuous-time quantum Monte Carlo and analytical methods, 
we show how 
the 
OSMP emerges as a stable dehybridization fixed point. Concomitantly,
the stability of the OSMP is demonstrated.
Connections of this mechanism with partial localization-delocalization transition 
in other strongly correlated metals are discussed.
\end{abstract}


\maketitle
{\it Introduction.~~} Iron-based 
 superconductors \cite{Kamihara2008,Johnston2011,Wang_Lee2011,Elbio_rmp,Dai_RMP2015,Si2016,Hirschfeld2016,Yi2017}
have several salient features in their normal state.
They are in proximity to antiferromangetic and other electronic orders, 
highlighting the role of collective spatial correlations.
They are bad metals, with the electrical resistivity
 at room temperature reaching the Mott-Ioffe-Regel limit, implicating the role of
strong electron correlations. Finally, they contain several active $3d$ orbitals whose degeneracies are 
 broken due to the crystalline environment. As such,
 they exemplify correlated multi-orbital metals. 
 Studies in such systems have revealed 
 an orbital-selective Mott phase (OSMP) \cite{Anisimov2002},
  in which the Mott-localized orbitals and metallic orbitals coexist. 
Indeed,
iron-based superconductors have been the setting in which orbital-selective correlations
 and proximity to OSMP
have been evidenced 
by angle-resolved photoemission spectroscopy (ARPES)~\cite{Huang2022,Yi2013,Yi2015,Liu2015}
and 
a variety of other experiments~\cite{Ding2014,Li2014,Gao2014}. 
Proximity to OSMP
serves to anchor the physics of  orbital selectivity in multi-orbital bad metals,
as evidenced by 
the observation of the high-energy Hubbard bands in the single-particle
spectrum of FeSe \cite{Watson2017,Evtushinsky2016} and FeTe$_{1-x}$Se$_{x}$ \cite{Huang2022}.
More generally, OSMP has been discussed in the ruthenates and a variety of other systems
\cite{Neupane2009,Kim2021.1,Mukherjee2016,Qiao2017}.

An OSMP is 
usually seen in models with the different $3d$ orbitals
 that do not kinetically
couple to each other \cite{Anisimov2002}. In the case of FeSCs, 
symmetry dictates the existence of inter-orbital hopping, 
which hybridizes between the
orbitals. Earlier analyses have concluded the stability of the OSMP against
such hybridization \cite{Yu2017,Yashar2017}. This conclusion, however, has recently been questioned \cite{Kugler2021}. 
In the meantime, experimentally, experimentally, hybridization has emerged as a tool
to probe the development of OSMP. In the FeTe$_{1-x}$Se$_{x}$ series, it has been shown that the 
renormalized 
hybridization between the most correlated $3d_{xy}$ orbital and the relatively weakly-correlated 
$3d_{z^2}$ systematically decreases as the doping $x$ decreases \cite{Huang2022}.
When $x$ approaches $0$, the
hybridization renormalizes towards zero, which is accompanied by a large reconstruction of the Fermi surface
from one that incorporates the $3d_{xy}$ states to one that does not. In this way, the inter-orbital hybridization
not only serves as a characterization of the approach towards OSMP, but also provides a means of using 
the less correlated electron states to probe the nature of the localizing orbitals.
In this way, the experiment motivates the notion that an OSMP is synonymous to correlation-driven dehybridization.

In this Letter, we develops this notion by identifying the OSMP with a dehybridizing fixed point. 
We interpret the slave-spin-based results \cite{Yu2017,Yashar2017} in terms of 
a competition between the hybridization and spatial correlations, which motivates the analysis of
effective models that arise from the extended dynamical mean-field theory (EDMFT). This is done
using a combination of continuous-time quantum Monte Carlo and analytical methods,
which demonstrate that a dehybridization fixed point characterizes the OSMP.
In turn, our results provide evidence that
the OSMP is a stable phase in the renormalization group sense.
Our results reveal and highlight an intriguing connection between the emergence of OSMP in
multi-orbital Hubbard models and the partial localization-delocalization transition that has been extensively discussed in
a variety of strongly correlated metals.

{\it Model and method.~~}
We consider 
a two-orbital Hubbard model on a square lattice with the following Hamiltonian \cite{Kugler2021}:
\be 
H_{\rm moh} = \sum_{ij,a,b,\sigma}d_{i,a,\sigma}^\dag( t_{ij,ab}-\mu \delta_{ij}\delta_{ab})d_{j,b,\sigma} +
\sum_{i} h_{i} \, .
\label{eq:ham}
\ee 
Here, $d_{i,a,\sigma}^\dag$ creates an electron at site $i$ in orbital $a$ with spin $\sigma$,
$t_{ij,ab}$ denotes a hopping matrix 
containing both intra-orbital and inter-orbital hopping parameters, and
 $\mu$ is the chemical potential.
 In addition,
 $h_i$ describes 
 local interactions with both the Hubbard interaction of strength $U$ and Hund's coupling of strength $J_H$. 
 We consider both the
Ising-anisotropic and ${\rm SU(2)}$-symmetric cases
(see Supplementary Materials).
The hopping matrix 
has the following form:
\ba 
\epsilon_k = 
\begin{bmatrix}
-2t_{11}(\cos k_x+\cos k_y)  & t_{12}(\cos k_x-\cos k_y) \\
t_{12}(\cos k_x-\cos k_y) & -2t_{22}(\cos k_x+\cos k_y)
\end{bmatrix} \nonumber 
\, ,
\\
\ea 
with $\epsilon_k$ the Fourier transformation of $t_{ij}$. $t_{11}(t_{22})$ is the nearest-neighbor intra-orbital hopping of orbital $1(2)$;
we work with $|t_{11}| < |t_{22}|$ so that, for the same interaction parameters of the two orbitals, orbital $1$ will be more strongly correlated.
In addition, $t_{12}$ is the nearest-neighbor inter-orbital hopping between two orbitals. 
The two orbitals 
are taken to be two different representations of the $C_4$ rotation, 
which implies the absence of any
onsite inter-orbital hybridization.

\begin{figure}[b!]
    \centering
    \includegraphics[angle=0,width=0.95\linewidth]{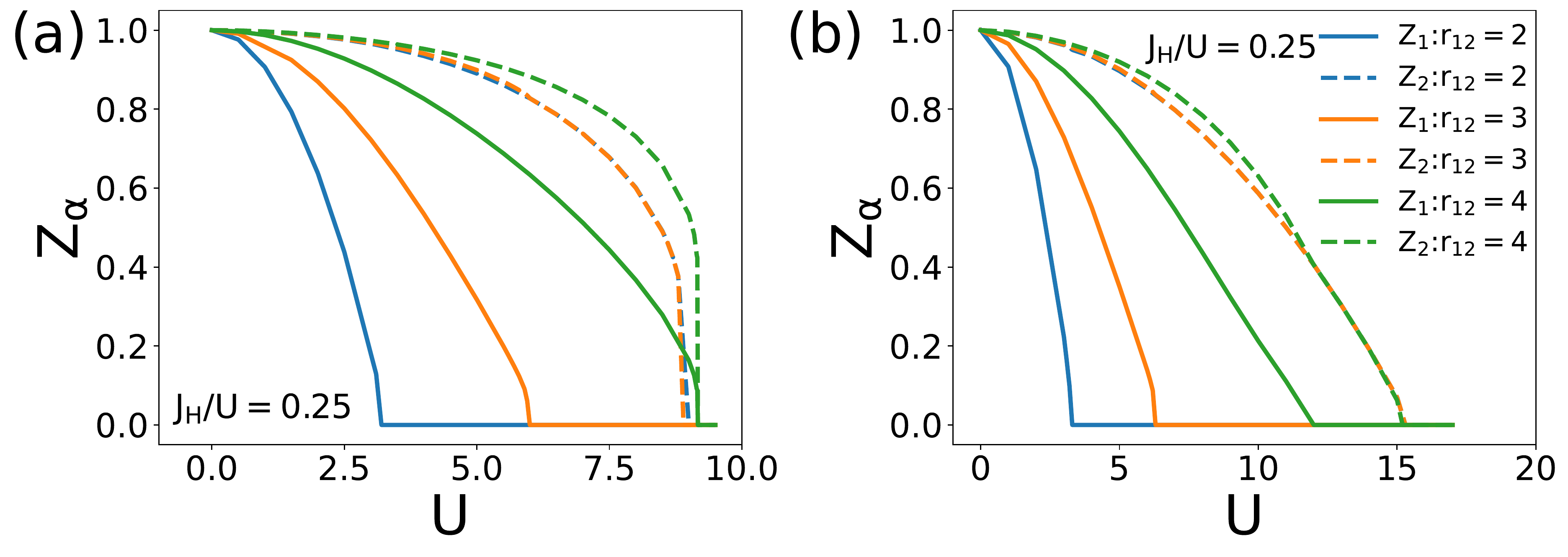}
    \caption{Quasiparticle weight vs. $U$, for fixed $J_H/U=0.25$, at half-filling and different values of $r_{12}=t_{12}/t_{11}$ in the Ising-anisotropic (a) and $SU(2)$-symmetric (b) two-orbital Hubbard models. }
    \label{fig:ss}
\end{figure}

{\it OSMP from the slave-spin approach.~~} 
The slave-spin method has been used in a variety of contexts.~\cite{Medici2005,Yu2012}.
Here, we utilize 
 the $U(1)$ slave spin approach to solve the model~\cite{Yu2012}
 and study
 both Ising-symmetric and $SU(2)$-symmetric models. For definiteness, we consider the hopping
 parameters
$t_{11}=0.5$ and $t_{22}=3$, with a fixed ratio for the Hund's coupling,
 $ J_H/U=0.25$ and at different values of $t_{12}$. 
 We take
  $U$ as the tuning parameter and track the quasi-particle weights $Z_a$ of the two orbitals.
A metallic (Mott-localized) orbital $a$ is 
implicated by the quasi-particle weight $Z_a  \ne 0 ( = 0)$.
As shown in Fig.~\ref{fig:ss}, in each case, there exists a nonzero parameter region of OSMP that is characterized by the 
Mott behavior of orbital $1$ ($Z_1=0$) and metallic behavior of  orbital  $2$ ($Z_2\ne 0$). 
We observe that the region of OSMP 
becomes smaller 
when  $t_{12}$ is increased, but does not disappear until
$t_{12}$ is above certain
threshold value. 

\begin{figure}[t!]
    \centering
    \includegraphics[width=0.45\textwidth]{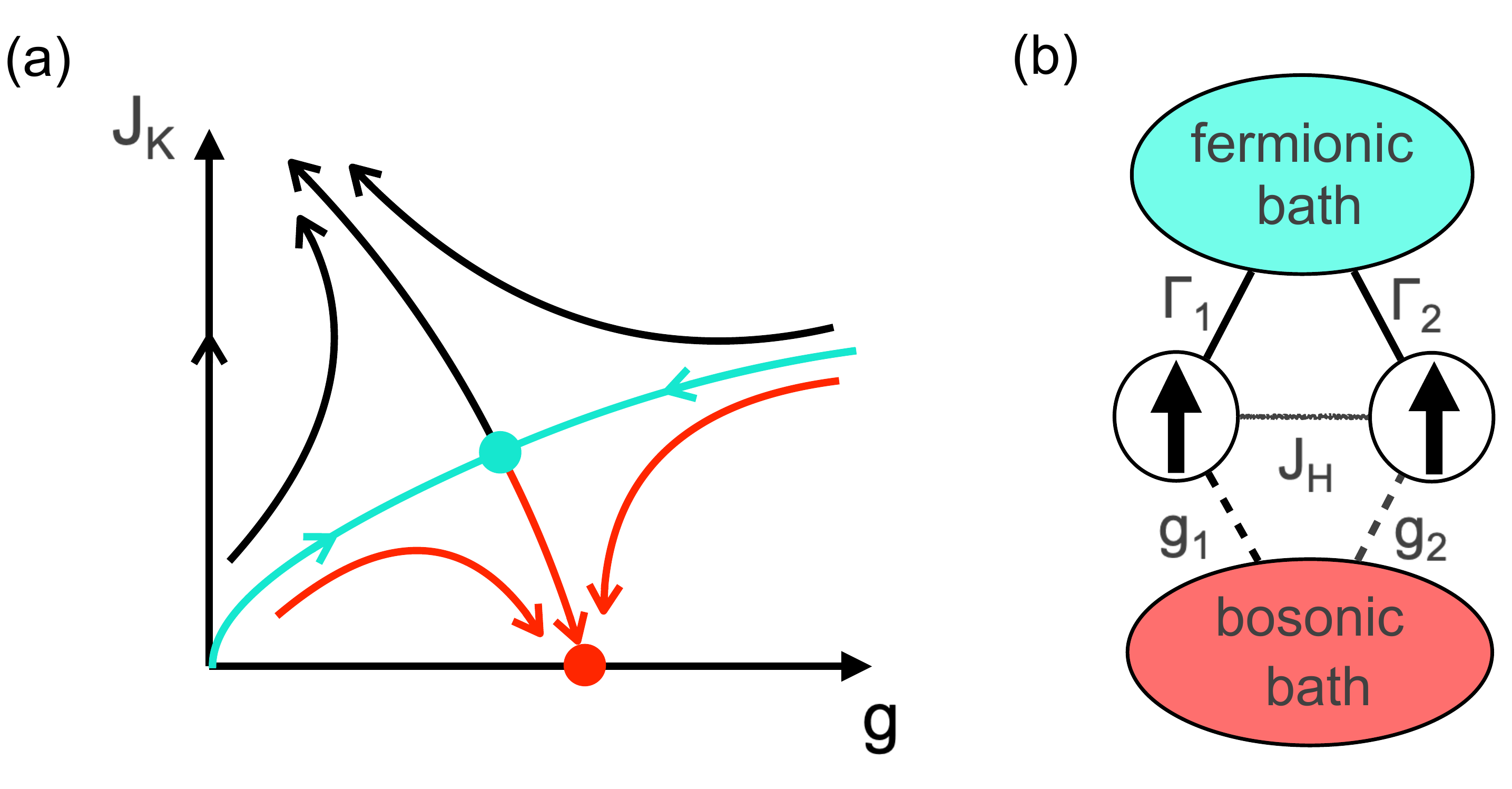}
    \caption{(a) Renormalization-group flow of a single-impurity Bose-Fermi Kondo model. $J_K(g)$ denotes the coupling between the impurity and  fermionic/bosonic bath. There are two stable fixed points: one is
    characterized by a relevant $J_K$, which corresponds to the Kondo-screened/hybridized phase; and the other 
    by an irrelevant $J_K$, which corresponds to the Kondo-destroyed/dehybridized phase.
    (b) Illustration of the Hund's-coupled two-orbital Bose-Fermi Anderson model that arises from the EDMFT approach of
    the two-orbital Hubbard model. }
    \label{fig:rg}
\end{figure}

In other words, OSMP in this two-orbital Hubbard model is stable
against inter-orbital hopping in the slave-spin approach
in consistency with the results 
for other multi-orbital models \cite{Yu2017,Yashar2017}.
Importantly, in the $U(1)$ slave-spin formulation, the spin degrees of freedom is captured by
 the fermions (the ``spinons"). 
 The dispersing of the spinons encodes
 dynamical spatial spin correlations,
 which compete against the hybridization. Our 
interpretation is that, this competition is responsible for 
 the stability of the OSMP in the slave-spin approach. 
 This is analogous to what happens in the self-consistent
  Bose-Fermi Kondo model where, as illustrated in 
  Fig.~\ref{fig:rg}(a),
   the spin-spin correlations suppress
  the Kondo effect and lead to a Kondo-destroyed fixed point 
  \cite{Zhu02.1,Zarand02}.
 This motivates a more direct and explicit analysis of the competition
between the dynamical spatial spin correlations and hybiridization.

{\it
Dynamical spatial spin correlations and EDMFT.~~} 
To incorporate the effect of dynamical spatial spin correlations, we treat the multi-orbital Hubbard model
in terms of EDMFT. In this approach, one generalizes the Hamiltonian to $H = H_{\rm moh} + H_J$
by introducing
 the following explicit
 intersite spin-spin interactions
\ba 
H_J &=& \sum_{a,\mu,\langle i,j\rangle }J^\mu_{a}S_{i,a}^\mu S_{j,a}^\mu +J^\mu_{12}\sum_{\langle i,j\rangle }(S_{i,1}^\mu S_{j,2}^\mu+S_{i,2}^\mu S_{j,1}^\mu )  \, ,
\nonumber \\
\label{eq:ham_spin}
\ea 
where $S^{x,y,z}_{i,a}=
d_{i,a}^\dag \sigma^{x,y,z}d_{i,a}/2$ are the spin operators of orbital $a$ and $\sigma^{x,y,z}$ are three Pauli matrices. $J^\mu_{1},J^\mu_2$ represent intra-orbital spin-spin interactions of two orbitals, and $J^\mu_{12}$ denotes inter-orbital spin-spin interaction. We have $J^\mu_{1,2,12}=J_{1,2,12}$ for the $SU(2)$-symmetric model and $J^\mu_{1,2,12}=J_{1,2,12}\delta_{\mu,z}$ for the Ising-anisotropic model. 
We expect that spin fluctuations have a larger contribution from the more correlated orbital $1$,  but the Hund's coupling would imply significant contributions from orbital $2$ as well.

After diagonalizing the matrix $J_{q,ab}$, we can rewrite Eq.~\ref{eq:ham_spin} via its eigenvalues and eigenvectors
\ba 
H_J  
=
\sum_{q,\mu}[ J^\mu_{+,q}S^\mu_{+,q}S^\mu_{+,-q} +J^\mu_{-,q}S^\mu_{-,q}S^\mu_{-,-q} ] \, .
\ea 
It's adequate to consider the most negative 
eigenvalues
and the corresponding eigenvectors:
\ba 
J^\mu_{q,- }&=&
\frac{1}{2}
\left [
J^\mu_{1}+J^\mu_2
-  
\sqrt{(J^\mu_1-J^\mu_2)^2+4(J^\mu_{12})^2}
\right ] \nonumber\\
&&[\cos(q_x)+\cos(q_y) ] \, , \nonumber \\
S_{-,q}^\mu &=& \cos(\theta/2)S_{1,q}^\mu +\sin(\theta/2)S_{2,q}^\mu \, ,
\ea 
where $\tan(\theta) = 2J_{12}/(J_1-J_2) $.

We now treat the model $H$
with EDMFT by mapping the lattice model to an effective two-orbital Bose-Fermi Anderson model (BFA model) \cite{PhysRevB.61.5184}:
\ba 
S_{\rm BFA}&=&\sum_{\omega,\sigma,a,b} d_{a,\sigma}^\dag[ (-i\omega-\mu)\delta_{ab}  +\Delta_{ab}(i\omega) ]d_{b,\sigma} \, ,
\nonumber \\
&&
-\frac{1}{2}\sum_{\Omega,\mu}{S}^\mu_{-}(i\Omega)\chi^{\mu,-1}_{0}(i\Omega) {S}^\mu_{-}(-i\Omega) +\int_\tau 
h_{0} \, .
\ea 
In the BFA model, two orbitals are coupled to fermionic baths denoted by $\Delta(i\omega)$ and bosonic baths denoted by $\chi^\mu_{0}(i\Omega)$. The two baths capture the effect of single-particle 
dynamics and 
dynamical spatial spin correslations, respectively~\cite{PhysRevB.61.5184}. 
Here, $h_0$ describes the local, Hubbard and Hund's, couplings. An illustration of the model is shown in Fig.~\ref{fig:rg}(b) where $g_{1}=\cos(\theta/2)g, g_2=\sin(\theta/2)g$ denote the effective coupling to the bosonic bath. The Hamiltonian formula of the model is shown in the Supplementary Materials. 

The bath functions $\Delta(i\omega)$ and $\chi^{\mu,-1}_0(i\Omega)$ are determined self-consistently via 
\ba 
G_{loc}(i\omega) 
&=&\sum_k[i\omega -\epsilon_k+\mu -\Sigma(i\omega)]^{-1} \nonumber\\
&=& 
[(i\omega +\mu -\Delta(i\omega))^{-1}-\Sigma(i\omega)]^{-1}
\nonumber \\
\chi^\mu_{loc}(i\Omega)
&=& \sum_q [J^\mu_{-,q} + M^\mu(i\Omega)]^{-1} 
\nonumber\\
&=& [{\chi^{\mu,-1}_0(i\Omega) +M^\mu(i\Omega)}]^{-1} \, .
\label{eq:self_cons}
\ea 
Here, $G_{loc}(i\omega),\chi^\mu_{loc}(i\Omega)$ are single-particle Green's function of two orbitals and spin-spin correlation functions of $S_-$ respectively. $\Sigma(i\omega),M(i\Omega)$ are the corresponding self-energy ~\cite{PhysRevB.61.5184}. $\Delta(i\omega)$, $G_{loc}(i\omega)$ and $\Sigma(i\omega)$ are all $2\times 2$ matrices in orbital space and block diagonal due to the symmetry: $\Delta(i\omega)=\text{diag}[\Delta_1(i\omega),\Delta_2(i\omega)]$, $G_{loc}(i\omega)=\text{diag}[G_1(i\omega),G_2(i\omega)]$, $\Sigma(i\omega)=\text{diag}[\Sigma_1(i\omega),\Sigma_2(i\omega)]$ \cite{Kugler2021}. $\chi^\mu_0(i\Omega)$, $\chi^\mu_{loc}(i\Omega)$ and $M^\mu(i\Omega)$ are complex numbers for each
$\mu$. In the Ising-anisotropic case, we only keep the $z$ component, and for the $SU(2)$-symmetric case, the
 three components are identical.

We turn next to analyzing the stability of OSMP via the above equations. 
In OMSP, the self-energy of
the more correlated orbital $1$ 
would diverge $\text{Im}[\Sigma_{22}(\omega\rightarrow 0)]\rightarrow \infty$ due to Mott behaviors. 
This gives a bath function $\Delta_{1} \sim \sum_k t_{12}^2(\cos(k_x)-\cos(k_y))^2G_{2}(i\omega,k)$,
showing that $\Delta_1$ is always gapless whenever the inter-orbital hopping is non-zero and the
 orbital $2$ is metallic \cite{Kugler2021}. 
However, in such a  case, as illustrated in
  Fig.~\ref{fig:rg}(a),  the spin-spin correlations compete against (in our case) hybridization and allows for a new stable fixed point
  where the Kondo effect is destroyed.
We now demonstrate that, for the two-orbital BFA models, this type of fixed point does emerge 
leading to 
an OSMP. In turn, the OSMP phase is stable against the inter-orbital hoping.

{\it The Ising-anisotropic model.~~}
Inspired by the previous analytical argument, we now solve the Ising-anisotropic BFA model via 
a continuous-time quantum Monte Carlo method \cite{Ang2019,Pixley2013,Otsuki2013,Gull2011}
and demonstrate 
the existence of a dehybridization fixed point and its correspondence to the OSMP phase.
Without loss of generality, we take two gapless fermionic baths and a sub-ohmic bosonic bath. The corresponding spectral functions are 
\ba 
&&\text{Im}[\Delta_1(\omega-i0^+)] = \Gamma_1 \theta(D-|\omega|) \nonumber \\
&&\text{Im}[\Delta_2(\omega-i0^+)] = \Gamma_2 \theta(D-|\omega|) \nonumber \\
&&\text{Im}[\chi^{-1}_0(\omega-i0^+)] = g^2\frac{1+s}{\Lambda^{1+s}}\omega^s\theta(\Lambda-\omega)\theta(\omega) \, .
\ea  
Here, $\theta(x)$ is the step function, $\Gamma_{a}$ is the hybridization strength between 
the orbital $a$ and fermionic bath $a$, $g$ is the coupling strength between $S^z_-$ and bosonic bath,
and $D,\Lambda$ are the bandwidths of the fermionic and bosonic baths respectively. 
Since, $\Delta_1\sim \sum_k t_{12}^2(\cos(k_x)-\cos(k_y))^2G_{22}(i\omega,k)$,
we let two fermionic baths to have the same bandwidth but different hybridization strengths, i.e. $\Gamma_1\ne \Gamma_2$. 
Finally, $s$ is the exponent that characterizes bosonic bath.
In the calculation, we set $U=0.04, J_H=0.2U, D=\Lambda=1,\Gamma_1=\frac{1}{4}\Gamma_2 =0.1$, and $s=0.8$. 
In addition, we pick $\tan(\theta/2) = \frac{1}{5}$ and $S^z_- = \frac{5}{\sqrt{26}}S^z_1 + \frac{1}{\sqrt{26}}S^z_2$, such that the spin dynamics are dominant by 
orbital $1$. We use $g$ as the tuning parameter.

\begin{figure}[t!]
    \centering
    \includegraphics[width=0.45\textwidth]{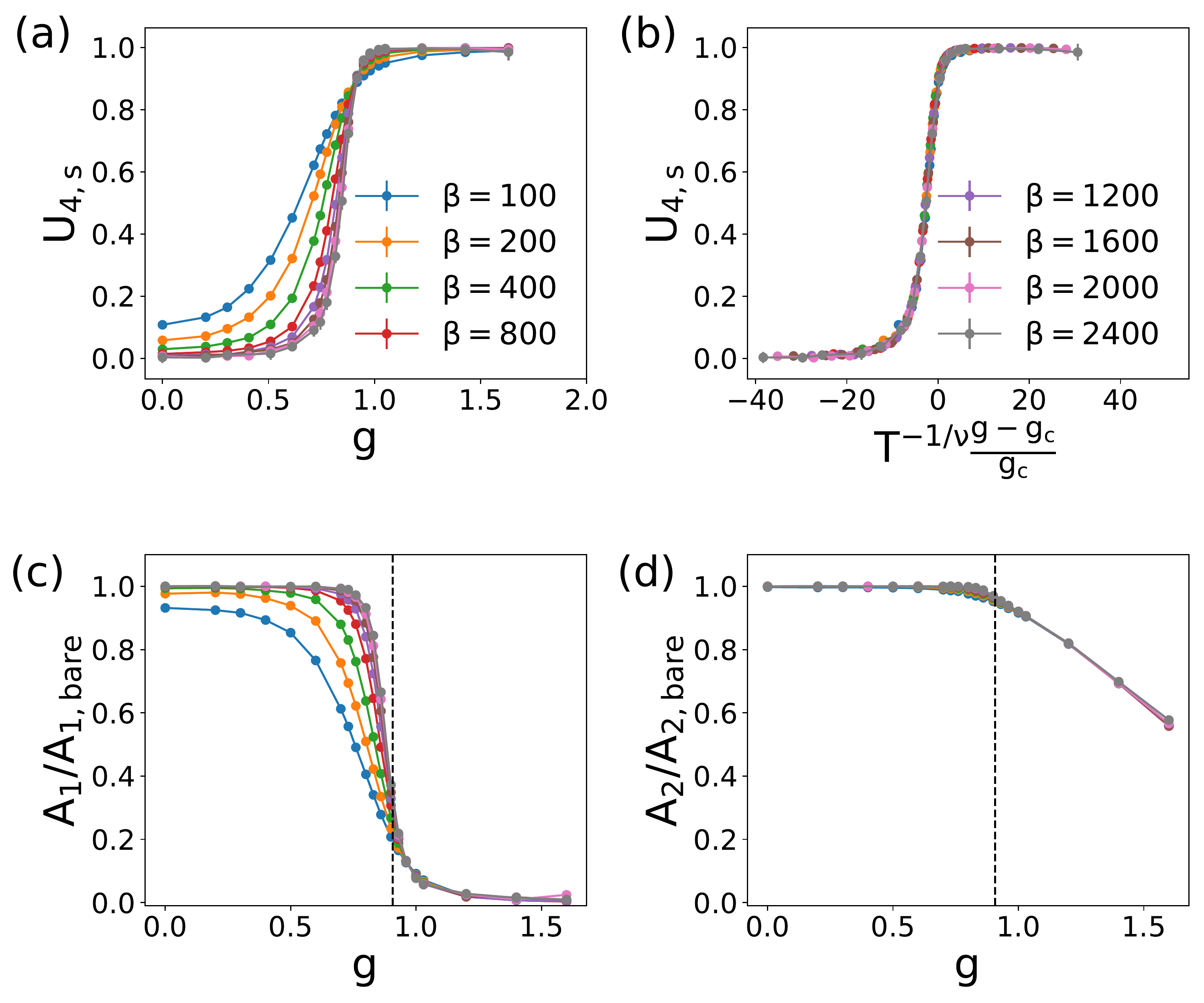}
    \caption{\label{fig:u4}
    (a) Binder cumulant $U_{4,s}(\beta,g)$ vs the bosonic coupling $g$. (b) Scaling collapse of the data optimized by $g_c/D = 0.90(1), \nu^{-1}=0.468(12)$. Evolution of $A_{a}$ of 
     orbital $1$ (c) and 
     orbital $2$ (d) at various temperatures. The dashed line marks the position of the QCP.}
\end{figure}

To detect the quantum phase transition, we track the evolution of a binder cumulant defined as 
\be 
U_{4,s} = \frac{1}{2}\frac{\langle (m_1+m_2)^4 \rangle}{\bigg(\langle (m_1+m_2)^2 \rangle\bigg)^2} \, ,
\ee
where $m_a = \frac{1}{\beta}\int_0^\beta S_a^z(\tau) d\tau$. In Fig.~\ref{fig:u4} (a) (b), we demonstrate the crossing behavior and the collapsing of the binder cumulant, which suggest a quantum phase transition \cite{Ang2019}. 
As we show later, the quantum critical point (QCP) at $g=g_c$ separates a stable fixed point describing the 
Fermi liquid phase at small $g$ and a separate stable fixed point corresponding to the OSMP phase at large $g$. 

To understand the nature of the two observed phases, 
we 
examine the single-particle 
density of states at zero energy $\rho_{a}(E=0)$. However, this requires the knowledge of real-time dynamics and the analytical continuation of the imaginary-time data is challenging. Instead, we consider the commonly used approximation: $\rho(E=0)\sim A_a=-\frac{\beta}{\pi} G_{a}(\tau=\frac{\beta}{2})$~\cite{dos_proxy}. $A_a$ is related to the density of states of orbital $a$, $\rho_{a}(\epsilon)$, via
\be 
A_a = -\beta G_{a}(\tau=\frac{\beta}{2}) = \int_{-\infty}^{\infty} \frac{\beta\rho_{a}(\epsilon)}{2\cosh(\beta\epsilon)}d\epsilon \, .
\ee
At zero temperature, $\beta\rightarrow \infty$, the kernel becomes a delta function and 
 $A_a$ is exactly the zero-energy density of states, $\rho_{aa}(\epsilon=0)$ and can be used to detect the
 delocalization-localization transition. 
 In Fig.~\ref{fig:u4} (c) (d), we show the evolution of $A_a$ normalized by its bare value $A_{a,bare}$ at $g=U=J_H=0$. When $g<g_c$, both $A_1,A_2$ remains large and suggests 
 that both orbitals are itinerant.
 When $g>g_c$, $A_2$ still acquire a large value $>50\%A_{2,bare}$. However, $A_1$ is highly suppressed and is smaller than 
 $0.1\% A_{1,bare}$ 
 (which is already at the same order of numerical 
 uncertainty) when $g\gtrsim 1.0$. This indicates the system
 is in the OSMP at $g>g_c$.

\begin{figure}[b!]
    \centering
    \includegraphics[width=0.45\textwidth]{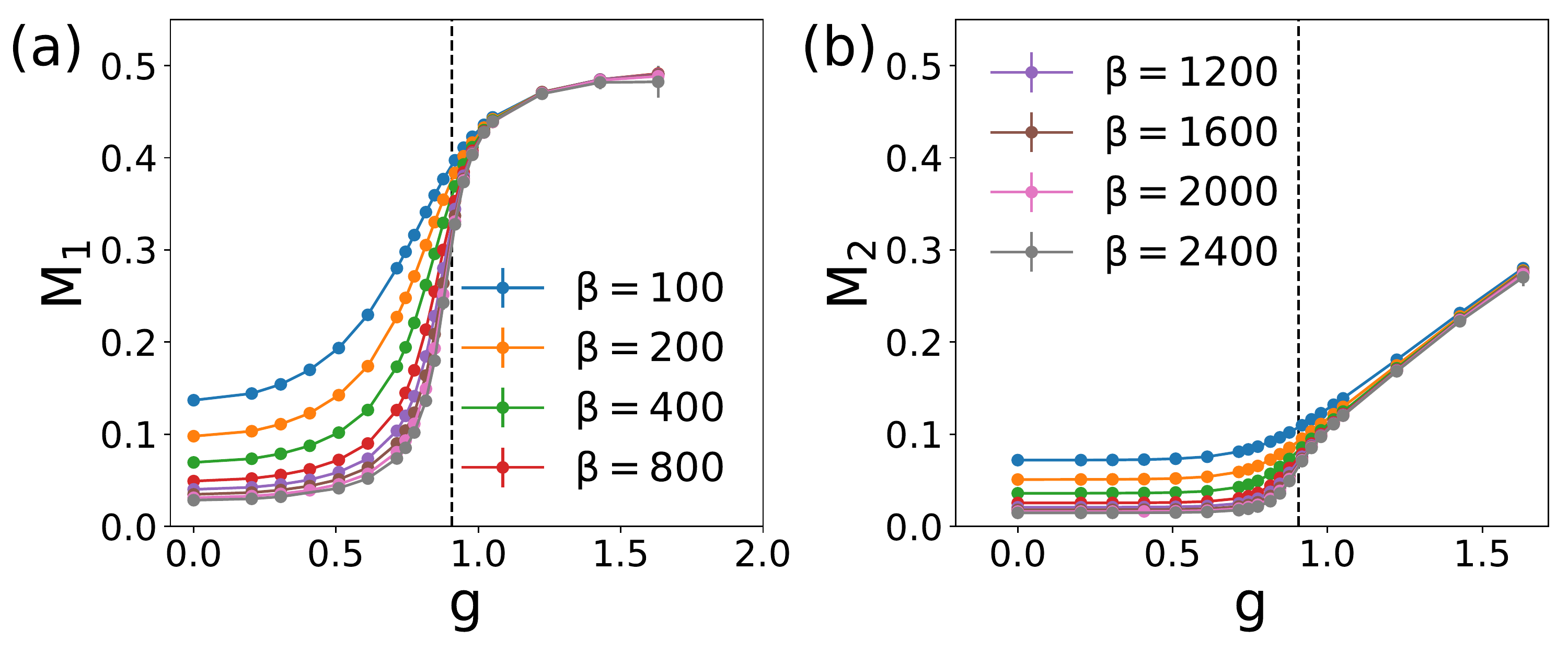}
    \caption{\label{fig:mag}
    $M_a$ of 
    the more strongly correlated orbital  $1$ (a) and
    the less correlated orbital $2$ (b) as a function of $g$ at various temperature. The dashed line marks the position of the QCP.}
\end{figure}

We also evaluate the root mean square of local magnetization of the two orbitals: 
\be 
M_a = \sqrt{\langle m_a^2\rangle } \, .
\ee
$M_a$ measures the size of local moment formed by orbital $a$. For the Mott localized orbital, a large $M_a$ develops and its value saturates to the maximum value $\frac{1}{2}$ when deep inside the localized phase. As illustrated in Fig.~\ref{fig:mag}, both $M_1,M_2$ are small when $g<g_c$ that is consistent with
both orbitals being itinerant. When $g>g_c$, $M_1$ quickly reaches the saturation value $1/2$ and $M_2$ remains small;
like the density of states, this implies the OSMP behavior.
 We also notice the
 more weakly-correlated orbital $2$ still has a sizable local moment formed at $g>g_c$. 
 This partially localizing behavior comes from the ferromagnetic correlations between the 
 two orbitals generated by the Hund's coupling and, effectively, by the bosonic bath as well. 
 However, $M_{2}$ is far from saturation,
 which is consistent with orbital $2$ remaining itinerant.

{\it The $SU(2)$-symmetric model.~~}
We next turn to the two-orbital BFA model with $SU(2)$ symmetry. Here, we
are able to 
analyze the model analytically at the saddle-point level. We first take the slave-spin approach
$d^\dag_{a,\sigma} =\tilde{S}^+_{a}f_{a,\sigma}^\dag$, where $\tilde{S}_a^+$
 is the U(1) slave-spin field and $f_{a,\sigma}$ is the slave-fermion field. We then introduce $z_a = \langle \tilde{S}^+_a\rangle$, $G^f_{ab}(\tau-\tau') = -\frac{1}{2}\sum_\sigma \langle f_{a,\sigma}(\tau)f^\dag_{b,\sigma}(\tau')\rangle $. The model can be solved at
 a  saddle-point level and we find the following OSMP solution
\ba 
&&z_1 =0,\quad z_2 \ne 0 \nonumber \\
&&G^f_{11}(\tau) \sim- \frac{\text{sgn}(\tau)}{|\tau|^{\frac{1-s}{2}}}\, ,\quad  G^f_{22}(\tau) \sim- \frac{\text{sgn}(\tau)}{|\tau|} \, ,
\ea 
As in the standard slave-particle method, $z_1=0$ and $z_2 \ne 0 $ 
corresponds to an OSMP phase 
with Mott-localized orbital $1$ and itinerant orbital $2$. Thus, we again realize
an OSMP phase as a dehybridization fixed point in the $SU(2)$-symmetric model. 

{\it Discussion.~~}
Our analysis has shown 
 that dynamical spatial spin correlations impede the inter-orbital hopping and allow for a dehybridization fixed point, which is tantamount to an OSMP phase. This finding provides a new lens to view prior works on the effect of the inter-orbital hopping.
We have already discussed that the U(1) slave-spin approach is able to access the OSMP fixed point. 
Such an OSMP phase may be considered 
as the U(1)-slave-spin analogue of the Z$_2$-slave-spin-derived ``orthogonal metal" \cite{Nandkishore2012};
as in the latter case, the OSMP phase corresponds to a distinct fixed point. 
In the same vein, EDMFT is able to access the
dehybridization fixed point of Fig.~\ref{fig:rg}(a) by treating the spatial correlations dynamically;
the latter is kept track of in the effective model via the bosonic bath.
This is to be contrasted to the
DMFT approach,
in which only the hybridization/Kondo process is treated dynamically [any magnetic order appears through a static (Hartree-Fock)
treatment of the spin-exchange interactions]. We propose this as underlying the DMFT result of 
Ref.~\onlinecite{Kugler2021}, which  found that the inter-orbital hopping destabilizes the OSMP of the zero-hopping model
and allows for only the hybridizing/Kondo fixed point.
Beyond the realm of multi-orbital Hubbard models, 
 the emergence of OSMP in our analysis resembles the development of Kondo destruction that 
 characterlizes the beyond-Landau quantum criticality of heavy fermion metals
 \cite{Gegenwart.08,Si2001,Coleman2001,Pepin2008}. Physically, both
 are examples of the general phenomenon of partial localization-delocalization transition 
 in the bad-metal regime of strongly correlated metals \cite{Pas21.1}.

{\it Conclusion.
~~}
We have addressed the issue of whether OSMP robustly develops within multi-orbital Hubbard 
models that 
contains inter-orbital hopping. An affirmative answer has been provided and, in the process,
the notion has been advanced that the OSMP is associated with a dehybridization fixed point. We have done 
so by analyzing the competition between the hybridization and spatial correlations in the multi-orbital
models. In particular, we carried out numerical and analytical analyses of 
 effective models that arise from the EDMFT formulation of the models. Our results directly connect
with recent ARPES results on the role of hybridization in the evolution of the iron 
chalcogenides towards an OSMP phase. More generally, our work reveals new connections
between the $3d$-electron-based bad metals in proximity to an OSMP with other categories 
of strongly correlated metals in near a (partial) localization-delocalization transition.

{\it Acknowledgements.~~}
We thank J. Huang and M. Yi  for useful discussions. 
Work at Rice was primarily supported by the 
U.S. Department of Energy, Office of Science,
Basic Energy Sciences, under Award No.\ DE-SC0018197
and additionally supported by
the Robert A.\ Welch Foundation Grant No.\ C-1411.
The majority of the computational calculations have been performed on the
Shared University Grid at Rice funded by NSF under Grant EIA-0216467, a partnership between
Rice University, Sun Microsystems, and Sigma Solutions, Inc., the Big-Data Private-Cloud 
Research Cyberinfrastructure MRI-award funded by NSF under Grant No. CNS-1338099 and by
Rice University, and the Extreme Science and Engineering Discovery Environment (XSEDE) by NSF
under Grant No. DMR160057. 
Work at Los Alamos was carried out under the auspices of the U.S. DOE
NNSA under Contract No. 89233218CNA000001; it was
supported by NNSA Advanced Simulation and Computing (ASC) Program and in part by the
Center for Integrated Nanotechnologies, a U.S. DOE BES
user facility.
 One of us (Q.S.) acknowledges the hospitality of the Aspen Center for Physics,
which is supported by NSF grant No. PHY-1607611.

\bibliography{osmp}

\newpage 

\onecolumngrid

\section{Supplementary Material}

\section{Model} 
The Hamiltonian of the $SU(2)$-symmetric two-orbital Hubbard model that we study via the slave-spin method is taken to be
\ba 
H_{\rm moh} &=& H_0+\sum_i h_i \nonumber \\
H_0 &=& \sum_{ij,a,b,\sigma}d_{i,a,\sigma}^\dag( t_{ij,ab}-\mu \delta_{ij}\delta_{ab})d_{j,b,\sigma} \nonumber \\
h_i &=&\bigg[U\sum_{a}n_{i,a,\up}n_{i,a,\dn}+(U'-J_H)\sum_\sigma n_{i,1,\sigma}n_{i,2,\sigma}+U'\sum_\sigma n_{i,1,\sigma}n_{i,2,\bar{\sigma}} \nonumber \\ 
&&-J_H\sum_\sigma(d_{i,1,\sigma}^\dag d_{i,1,\bar{\sigma}}d_{i,2,\bar{\sigma}}^\dag d_{i,2,\sigma}+d_{i,1,\sigma}^\dag d_{i,1,\bar{\sigma}}^\dag d_{i,2,\sigma}d_{i,2,\bar{\sigma}})\bigg]\nonumber,
\ea 
where $n_{i,a,\sigma}=d_{i,a,\sigma}^\dag d_{i,a,\sigma} $ is the density operator of orbital $a$ and spin $\sigma$ at site $i$. 
$U$ is the Hubbard interactions, $J_H$ the Hund's coupling and $U'=U-2J_H$. The interaction term can be written in a more compact 
form
\ba 
h_i
&=& \bigg[(\frac{U}{2}-\frac{5J_H}{8})(n_{i,1}+n_{i,2}-2)^2-\frac{3J_H}{8}(n_{i,1}-n_{i,2})^2 \nonumber\\
&&-2J_H \bm{S}_{i,1}\cdot \bm{S}_{i,2} -J_H\sum_\sigma d_{i,1,\sigma}^\dag d_{i,1,\bar{\sigma}}^\dag d_{i,2,\sigma}d_{i,2,\bar{\sigma}}
+(\frac{3U}{2}-
\frac{5J_H}{2})\sum_a n_{i,a}-(2U-
\frac{5J_H}{2}
)\bigg],
\ea 
where we have separate interactions into 
formally a density-density interaction (the first term), an orbital-orbital interaction (the second term), a spin-spin interaction (the third term),
a pair hopping term (the fourth term), a chemical-potential shift (the fifth term) and a constant.
 In practice, we shift the chemical potential $\mu \rightarrow \mu +\frac{3U-5J_H/2}{2}$, such that we have half-filling at $\mu=0$.

In the Ising-anisotropic case, we drop the spin-flipping and pairing hopping terms and have
\ba 
h_{i, {\rm Ising}}&=&
\bigg[U\sum_{a}n_{i,a,\up}n_{i,a,\dn}+(U'-J_H)\sum_\sigma n_{i,1,\sigma}n_{i,2,\sigma}+U'\sum_\sigma n_{i,1,\sigma}n_{i,2,\bar{\sigma}}\bigg] 
\ea 

The $t$-$U$-$J_H$-$J$ model we study via EDMFT has additional spin-spin interaction terms \cite{PhysRevB.61.5184}
with Hamiltonian: $H=
H_{\rm moh}+H_J$.
Through EDMFT, we map the lattice model to a 
two-orbital BFA
model 
\ba 
H_{\rm BFA} &=&h_0+H_c +H_{\phi} \nonumber \\
H_c &=&\sum_{k,a,\sigma}  \epsilon_{k,a}c_{k,a,\sigma}^\dag c_{k,a,\sigma} +\sum_{a,k,\sigma}V_{a}(d_{a,\sigma}^\dag c_{k,a,\sigma}+\text{h.c.})\nonumber \\
H_\phi &=&\sum_{q,\phi}\omega_q \phi^{\mu,\dag}_{q} \phi^\mu_q + \sum_{q,\phi} g^\mu S^\mu_-(\phi^\mu_{q}+\phi^{\mu,\dag}_{-q}).
\ea 
Here, 
$h_0$ is $h_i$ at a given site $i=0$, describing the Hubbard and Hund's interactions of the two orbitals.
$H_c\,(H_{\phi})$ contains the kinetic energy of the fermionic (bosonic) bath and the coupling between the baths and the two orbitals 
of $h_0$.
 We have $g^\mu = g\delta_{\mu,z}$ for the Ising-anisotropic model and $g^\mu = g$ for the $SU(2)$-symmetric model. 
 By integrating out fermionic baths $c_{k,a,\sigma}$ and bosonic baths $\phi_{q}^\mu$, we reach the action shown in the main text with 
\ba 
\Delta_a(i\omega) = \sum_{k}\frac{V_a^2}{i\omega-\epsilon_{k,a}}\quad\quad \chi_0^{\mu,-1}(i\Omega) =(g^\mu)^2 \sum_{q}\frac{2\omega_q}{\omega_q^2-\Omega^2}
\ea

\section{Supplementary data}

\begin{figure}[h]
    \centering
    \includegraphics[width=0.5\textwidth]{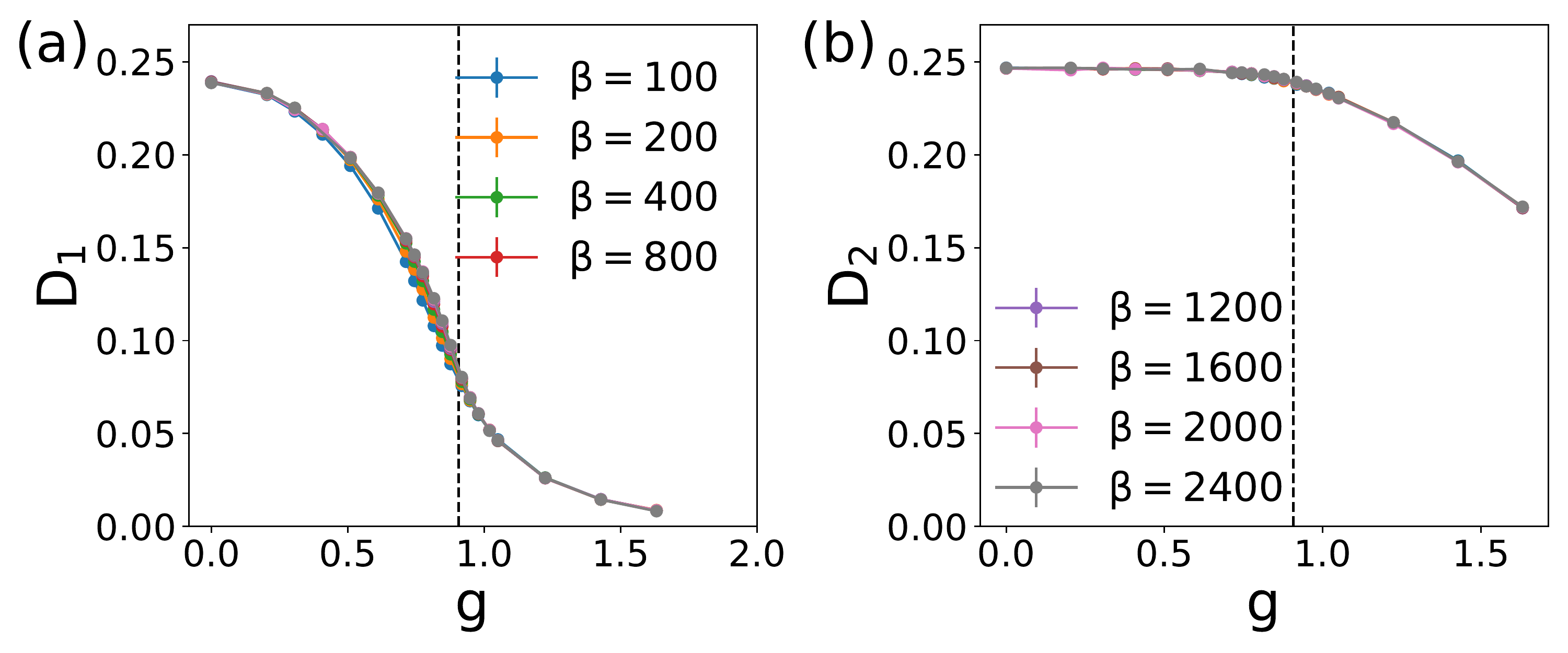}
    \caption{Doublon density of orbital 1 (a) and orbital 2 (b). The dashed line marks the QCP.
    \label{fig:doublon}}
\end{figure}

In Fig.\ref{fig:doublon}, we plot the evolution of the doublon density. 
We can observe a rapid suppression of the doublon density of orbital $1$ across the QCP,
which is consistent  with this orbital undergoing a Mott transition.

\section{Saddle-point analysis of the $SU(2)$-symmetric model}
In this section, we analyze the stability of OSMP in the $SU(2)$-symmetric model at the
saddle-point level. We take the following two-orbital BFA model
\ba 
S_{\rm BFA}
&=&\sum_{\omega,\sigma,a,a} d_{a,\sigma}^\dag\, [-i\omega-\mu  +\Delta_{a}(i\omega) ]\,d_{a,\sigma} \nonumber \\
&&
-\frac{1}{2}\sum_{i\Omega}{S}^\mu_{-}(i\Omega)\,\chi^{-1}_{0}(i\Omega) \,{S}^\mu_{-}(-i\Omega) +\int_\tau h'_0.\\
h'_0 &= &(\frac{U}{2}-\frac{5J_H}{8})(n_{1}+n_{2}-2)^2 - \frac{3J_H}{8}(n_1-n_2)^2
\ea 
The bosonic bath $\chi_0^{-1}$ dynamically generates an effective ferromagnetic spin-spin interaction term between the two orbitals, so we drop the corresponding part in $h'_0$. In addition, we also ignore the pair hopping term that is not important to the OSMP physics, and only keep the local interactions in the formally density and orbital channels. Finally, we take the same bath functions as the Ising-anisotropic model, which has been given in the main text.

We use the $U(1)$ slave-spin representation to treat the model. Here, we take a spin-1 slave-spin in order to fully decouple the charge and spin degrees of freedom~\cite{spin1_slavespin}. The electron operators are written as the product of the slave-spin and slave-fermion operators: $d^\dag_{a,\sigma} =\tilde{S}_a^+ f^\dag_{a,\sigma}$. The corresponding physical Hilbert space is defined as follows
\ba 
&&|0\rangle_{d} = |-1\rangle_{\tilde{S}} |0\rangle_{f} \nonumber\\
&&|\up/\dn\rangle_{d} = |0\rangle_{\tilde{S}} |\up/\dn\rangle_{f} \nonumber\\
&&|\up\dn \rangle_{d} = |1\rangle_{\tilde{S}} |\up\dn \rangle_{f}  \, ,
\ea 
where $|0\rangle_{d/f},|\up/\dn \rangle_{d/f}, |\up\dn \rangle_{d/f}$ are the empty, singly-occupied and doubly-occupied states 
of the electron/slave-fermion operators. In addition, $|-1,0,1\rangle_{\tilde{S}}$ are the eigenstates of the slave-spin operator, $\tilde{S}^z$. 
The corresponding local constraints are $\tilde{S}_a^z+1=\sum_{\sigma} f_{a,\sigma}^\dag f_{a,\sigma}$. In the slave-spin representation, we rewrite the action as
\ba 
S &= &\int_\tau \sum_{\sigma,a}f_{a,\sigma}^\dag[\partial_\tau-\mu-i\lambda_a ]f_{a,\sigma}+S_{Berry}[\tilde{S}^\mu]  \nonumber \\
&&+\int_\tau i\sum_a\lambda_a(\tau)(\tilde{S}^z_a(\tau)+1)+\int_\tau H_{I,loc}'[\tilde{S}^z(\tau)]\nonumber \\
&&+  \int_{\tau,\tau'}\sum_{a,\sigma}\tilde{S}_a^+(\tau)f_{a,\sigma}^\dag(\tau)\Delta_a(\tau-\tau')\tilde{S}_a^-(\tau')f_{a,\sigma'}(\tau')\nonumber \\
&&-\frac{1}{2}\int_\tau \sum_{a,b,\mu}g_{a}g_b \frac{f_a^\dag (\tau)\sigma^\mu f_a(\tau)}{2}\chi_0^{-1}(\tau-\tau')\frac{f_b^\dag (\tau')\sigma^\mu f_b(\tau')}{2}\nonumber \, .\\
\ea
Here, $S_{Berry}$ is the Berry phase 
contribution
 of the slave spin; $\lambda_a$ is the Lagrange multiplier that enforces the constraint;
$g_{1}=\cos(\theta/2)$ and $g_2=\sin(\theta/2)$; and $\sigma^{x,y,z}$ are the three Pauli matrices. 
We then insert the following identities to the partition functions
\ba 
1&=&\int_{\eta,z}e^{i\eta_a(\tau)(z^\dag_a(\tau)-\tilde{S}_a^+(\tau))} \nonumber \\
&=&\int_{\widetilde{\Sigma}_{ba},G^f_{ab}}e^{i\widetilde{\Sigma}_{ba}(\tau'-\tau)\bigg[ G^f_{ab}(\tau-\tau')+\frac{1}{2}\sum_\sigma f_{a,\sigma}(\tau)f_{b,\sigma}^\dag(\tau') \bigg]}.
\ea 
We then have
\ba 
Z &=& \int_{f,f^\dag,\tilde{S},\lambda, G^f,\widetilde{\Sigma},\eta,z}
\exp(-S_{eff})\nonumber \\
S_{eff} &=&i \int_{\tau} \sum_{a,b}\bigg[-\widetilde{\Sigma}_{ba}(-\tau) G^f_{ab}(\tau)-\delta_{a,b}\eta_a(\tau)z^\dag_a(\tau) \bigg] +\int_{\tau,\tau'}\sum_{a,b,\sigma}f_{a,\sigma}^\dag(\tau)[\delta_{\tau,\tau'}\delta_{ab}(\partial_{\tau'}-\mu-i\lambda_a)+i\widetilde{\Sigma}_{ab}(\tau-\tau')]f_{b,\sigma}(\tau') \nonumber \\
&&+S_{Berry}[\tilde{S}] +\int_\tau\sum_a i\lambda_a(\tau)(\tilde{S}^z_a(\tau)+1)+\int_\tau H'_{I,loc}[\tilde{S}^z] + \int_\tau -i\sum_a\eta_a(\tau)\tilde{S}_a^+(\tau)\nonumber \\
&&+\int_{\tau,\tau'}\sum_{a,\sigma}z_a^\dag(\tau)f_{a,\sigma}^\dag(\tau)\Delta_a(\tau-\tau')z_a(\tau')f_{a,\sigma'}(\tau')\nonumber \\
&&+\int_{\tau,\tau'}g_ag_b\chi_0^{-1}(\tau-\tau')\bigg[ G^f_{ba}((\tau'-\tau)G^f_{ab}(\tau-\tau')\bigg]. 
\ea 
Note that in the final line, we keep the term that survives in the large $N$ limit when we generalize $SU(2)$ to $SU(N)$. As demonstrated in Ref.~\cite{largeN}, such a term would be the most important to realize a Kondo-destruction behavior or, equivalently,
the OSMP behavior.
 In addition, at the saddle point level,
 $z_a =\langle \tilde{S}_a^x\rangle$, and $G^f_{ab}(\tau)=-\frac{1}{2}\langle\sum_\sigma f_{a,\sigma}(\tau)f^\dag_{b,\sigma}(0)\rangle$.

We set $\lambda_a(\tau)=\mu=0$ to stay at half-filling and integrate out the slave-spin and $\eta$ fields
\ba 
\int_{\tilde{S},\eta}\exp\bigg\{ -S_{Berry}[\tilde{S}] -\int_\tau H'_{I,loc}[\tilde{S}^z] -\int_\tau -i\sum_a\eta_a(\tau)(S_a^\dag(\tau)-z^\dag_a(\tau))\bigg\} \approx  \exp\bigg(-\int_\tau \sum_a \frac{C}{8} |z_a(\tau)|^2\bigg) \, ,
\ea 
where $C$ is approximately the charge gap in the atomic limit. Since we treat the $z_a$ as a static field at the
 saddle-point level, the dynamical part of the $z_a$ field has been dropped. 
 In addition, we only keep the bilinear term and ignore the high-order contributions [such as $(z^\dag z)^2$]. 
 Due to the fact that $z_a = \langle \tilde{S}^+\rangle$ and the $U(1)$ gauge redundancy in the slave spin representation,
  we can treat $z_a$ as a real field. Furthermore, at the saddle point, we take static $z$ fields [$z_a(\tau)=z_a$]
   and let $i\tilde{\Sigma}_{ab}(\tau) = \Sigma_{ab}(\tau)$. This gives the following saddle-point equations 
\ba 
&&[G^{f,-1}(i\omega)]_{ab} = \begin{bmatrix}
i\omega  - z_1^2 \Delta_1(i\omega) - \Sigma_{11}(i\omega) & -\Sigma_{12}(i\omega) \\
-\Sigma_{21}(i\omega) & i\omega -z_2^2\Delta_2(i\omega)-\Sigma_{22}(i\omega)
\end{bmatrix}_{ab} \nonumber \\
&&\int_\tau z_a\Delta_a(\tau)G^f_{aa}(-\tau) +\frac{z_aC}{4}=0 \nonumber \\
&&\Sigma_{ab}(\tau) =2g_ag_b\chi_0^{-1}(\tau)G^f_{ba}(\tau)  
\label{eq:saddle_point}
\ea 
Equivalent to the standard slave-spin method \cite{Yu2012}, if $z_a\ne 0 \,(z_a=0)$, orbital $a$ would be metallic (Mott-localized). Then, an OSMP phase solution with the following formula 
\ba 
&&z_1 =0,z_2\ne 0 \nonumber\\
&&G^f_{11}(\tau) \sim- \frac{\text{sgn}(\tau)}{|\tau|^{\frac{1-s}{2}}} \nonumber \\
&&G^f_{22}(\tau) \sim- \frac{\text{sgn}(\tau)}{|\tau|}\nonumber \\ 
&&G^f_{12,21}(\tau) \sim 0. 
\label{eq:osmp_sol}
\ea 
satisfies the self-consistent equations \cite{largeN}. 
Note that a non-zero $G^f_{12}(\tau)$ would generate antiferromagnetic correlations between the two orbitals which can be seen from $\langle S_1(\tau)S_2(0)\rangle \sim G^f_{12}(\tau)^2$. Thus $G^f_{12}(\tau)$ is energetically disfavored by the (effective) Hund's coupling 
and goes to zero at the saddle point. Now, we prove Eq.~\ref{eq:osmp_sol} indeed satisfy the saddle point equations. From Eq.~\ref{eq:osmp_sol} and the third line of Eq.~\ref{eq:saddle_point}, 
\ba 
&&\Sigma_{11}(i\omega) \sim (i\omega)^{(1+s)/2} \nonumber  \\
&&\Sigma_{22}(i\omega) \sim (i\omega)^{1+s}\nonumber \\
&&\Sigma_{12}(i\omega)=\Sigma_{21}(i\omega)= 0
\ea 
Then from the first line of Eq.~\ref{eq:saddle_point}, 
\ba 
&&G^f_{11}(i\omega)^{-1} \sim i\omega - C_2(i\omega)^{{1+s}/2} \nonumber\\
&&G^f_{22}(i\omega)^{-1} \sim i\omega - iz_2^2 \Gamma_2 +C_2(i \omega)^{{1+s}/2} \nonumber \\
&&G^f_{12}(i\omega)=G^f_{21}(i\omega)=0.
\ea
Taking the
dominant-order contribution (note that $0\le s\le 1$) and transforming to the imaginary-time domain, we have
\ba 
&&G^f_{11}(\tau) \sim- \frac{\text{sgn}(\tau)}{|\tau|^{\frac{1-s}{2}}} \nonumber \\
&&G^f_{22}(\tau) \sim- \frac{\text{sgn}(\tau)}{|\tau|}\nonumber \\ 
&&G^f_{12,21}(\tau) = 0
\ea 
which are consistent with Eq.~\ref{eq:osmp_sol}. As for the second line of Eq.~\ref{eq:osmp_sol}, it can be satisfied by letting 
$z_1=0$ and $z_2 \sim \frac{D}{\Gamma_2}e^{-C/(4\Gamma_2)}\ne 0$. In conclusion, Eq.~\ref{eq:osmp_sol} indeed solves Eq.~\ref{eq:saddle_point} in the low-energy limit.

Alternatively, we can also analyze the effective free energy
\ba 
\beta F_{eff}[z_a,G^f_{ab}] &=& 
\beta \frac{C}{8}\sum_a z_a^2 -\int_\tau \sum_{a,b}g_ag_b\chi_0^{-1}(\tau)G^f_{ba}(-\tau)G^f_{ab}(\tau)\nonumber \\
&&-\text{Tr}\log
\begin{bmatrix}
-\partial_\tau - z_1^2 \Delta_1(\tau)- 2g_1^2\chi_0^{-1}(\tau)G^f_{11}(\tau) & -2g_{1}g_2\chi_0^{-1}(\tau)G^f_{21}(\tau) \\
-2g_{1}g_2\chi_0^{-1}(\tau)G^f_{12}(\tau) & -\partial_\tau -z_2^2\Delta_2(\tau)-2g_2^2\chi_0^{-1}(\tau)G^f_{22}(\tau)
\end{bmatrix}\nonumber .\\
\ea  
To see the existence of the OSMP solution, we extract the mass term of $z_a$. To do so, we define $G^f_{ab}[z_a]$ as the saddle point value of $G_{ab}^f$ at $z_a$. We can extract the mass term of $z$ via 
\ba 
m_a &= &\frac{\delta^2 F_{eff}[z_a,G_{ab}^f[z_a]]}{\delta z_a \delta z_a}\bigg|_{z_1=z_2=0} \nonumber \\
&\approx& \frac{C}{4}+\int \Delta_a(\tau)G^f_{aa}[z_a=0](-\tau)d\tau. 
\ea 
As discussed before, at $z_a=0$, we have $G^f_{aa}[z_a=0]\sim \frac{-\text{sign}(\tau)}{|\tau|^{(1-s)/2}}$. Approximately, its density of states in the low-energy limit is $\frac{1}{\pi}\text{Im}[G^f_{aa}[z_a=0]]=\frac{1+s}{4\Lambda_a^{(1+s)/2}}|\epsilon|^{-(s+1)/2}\theta(\Lambda_a-|\epsilon|)$,
where $\Lambda_a$ is the energy scale below which the spin dynamics of orbital $a$ are controlled by the bosonic bath. 
When $g_a$ increases, we expect $\Lambda_a$ also increases. Then we finally have 
\ba 
m_a \approx \frac{C}{4} - \frac{\Gamma_a}{1-s}\bigg[2+(1-s)\log(D/\Lambda_a)\bigg],
\ea
where we have assume the cutoff $\Lambda_a<D$ in the derivation. The Mott transition in orbital $a$ happens at 
$C= C_a = \frac{4\Gamma_a}{1-s}\bigg[2+(1-s)\log(D/\Lambda_a)\bigg]$, where $m_a=0$. 
Since $\Gamma_1<\Gamma_2,\Lambda_2>\Lambda_1$, $C_{1}<C_{2}$, we can tune $C$ (equivalently, tuning the 
Hubbard
 interaction $U$) to satisfy $C_{2}>C>C_{1}$. In this parameter region, 
 $m_1>0 $ and $m_2>0$. Then $z_1$ is gapped and does not condense, but $z_2$ condenses and acquires a non-zero expectation value. 
 This is exactly the OSMP solution.


\end{document}